\documentclass[conference]{IEEEtran}
\IEEEoverridecommandlockouts
\usepackage{cite}
\usepackage{amsmath,amssymb,amsfonts}
\usepackage{algorithmic}
\usepackage{graphicx}
\usepackage{textcomp}
\usepackage{xcolor}
\usepackage{booktabs}

\usepackage{subfigure}
\usepackage{float}

\usepackage{stfloats}

\usepackage{verbatim}
\usepackage{epstopdf}

\def\BibTeX{{\rm B\kern-.05em{\sc i\kern-.025em b}\kern-.08em
    T\kern-.1667em\lower.7ex\hbox{E}\kern-.125emX}}
\begin{document}

\title{Blind Channel Estimation for MIMO Systems via Variational Inference \\

}

\author{\IEEEauthorblockN{Jiancheng Tang, Qianqian Yang\IEEEauthorrefmark{2}, Zhaoyang Zhang}
\IEEEauthorblockA{College of Information Science and Electronic Engineering, Zhejiang University, Hangzhou 310007, China\\
Email: \{jianchengtang,qianqianyang20\IEEEauthorrefmark{2},ning\_ming\}@zju.edu.cn
}
\thanks{{\thefootnote}{*}This work is partly supported by the SUTD-ZJU IDEA Grant (SUTD-ZJU (VP) 202102), and partly by the Fundamental Research Funds for the Central Universities under Grant 2021FZZX001-20.}
}

\maketitle

\begin{abstract}
In this paper, we investigate the blind channel estimation problem for MIMO systems under Rayleigh fading channel. Conventional MIMO communication techniques require transmitting a considerable amount of training symbols as pilots in each data block to obtain the channel state information (CSI) such that the transmitted signals can be successfully recovered. However, the pilot overhead and contamination become a bottleneck for the practical application of MIMO systems with the increase of the number of antennas. To overcome this obstacle, we propose a blind channel estimation framework, where we introduce an auxiliary posterior distribution of CSI and the transmitted signals given the received signals to derive a lower bound to the intractable likelihood function of the received signal. Meanwhile, we generate this auxiliary distribution by a neural network based variational inference framework, which is trained by maximizing the lower bound. The optimal auxiliary distribution which approaches real prior distribution is then leveraged to obtain the maximum a posterior (MAP) estimation of channel matrix and transmitted data. The simulation results demonstrate that the performance of the proposed blind channel estimation method closely approaches that of the conventional pilot-aided methods in terms of the channel estimation error and symbol error rate (SER) of the detected signals even without the help of pilots.
\end{abstract}


\section{Introduction}
With the capability of substantially improving the channel capacity and spectrum efficiency, multi-input multi-output (MIMO) approach has been recognized as one of the key enabling techniques for reliable high data rate transmission\cite{b1},\cite{b2}. And the orthogonal frequency division multiplexing (OFDM) approach is deployed in MIMO systems to eliminate the intersymbol interference (ISI) resulted form the frequency-selective fading\cite{b3}  by breaking the broadband system into many different narrowband subchannels. However, the transmitted signals in subchannels still experience power attenuation caused by flat fading which impedes the reliable detection of received signals. Therefore the accurate estimation of channel state information (CSI) is essential to compensate the channel responses and recover the transmitted information\cite{b4}.



The channel estimation techniques for MIMO systems have been studied extensively in the past \cite{b5},\cite{b6}, and the existing techniques can be summarized in two categories: pilot-aided estimation and blind estimation. The former approach requires to transmit orthogonal pilot sequences in order to estimate the channel matrices, which occupies a considerable portion of spectrum resources thus reduces spectrum efficiency \cite{b7}. Furthermore, the overlap of frequency by adjacent cells in cellular system impedes the orthogonality of pilot, which is referred to as pilot contamination\cite{b8}. On the other hand, the blind estimation methods are studied to overcome the pilot contamination problem and reduce the spectrum overhead. The second-order statistics of received signals is utilized to derive the blind channel estimation algorithm in \cite{b9}, where the covariance matrix of the received signal is divided into an orthogonal signal subspace and a noise subspace by singular value decomposition (SVD). However the SVD operation of large-scale matrix works in an iterative manner, which result in prohibitive computational cost \cite{b10}.
 
Recently, deep learning (DL) based channel estimation methods have achieved promising results and attracted a lot of research interests \cite{b11}. In \cite{b12},\cite{b13}, the time-frequency response of pilots going through fading channels is regarded as a low-resolution image, which is then enhanced by a DL-based image super-resolution (SR) network to estimate the CSI. These works are designed for single user scenarios and still require a certain amount of pilots to initialize the time-frequency response images. In \cite{b14}, a blind channel estimator based on a denoising convolutional neural network (DnCNN) is proposed for massive MIMO systems, where the DnCNN is employed to remove the residual noise effects which cannot be averaged out through channel asymptotic orthogonality. However, the performance of this method degrades in limited-scale MIMO systems as the asymptotic orthogonality of the channel vectors only exists in large scale MIMO systems.

In this paper, inspired by the widely applied approximate inference algorithms, i.e., variational inference (VI) which could provide a good approximation to the complex distribution\cite{b15}, we propose a VI based blind channel estimation technique for MIMO systems. The conventional signal detection method exploits the maximum a posterior (MAP) receiver to perform the exact inference of transmitted signals, which requires the posterior probability of the CSI. However, this may be intractable due to the dimensionality of the channel matrix. To deal with this challenge, we utilize a neural network based VI framework to generate an auxiliary distribution, which is optimized by maximizing a lower bound to the log-likelihood of the received signal. The derived auxiliary distribution approximates the real distribution, from which, the estimation of the CSI can be sampled, and thus the transmitted signals can be detected using maximum likelihood estimation. The numerical results validate the effectiveness of the proposed blind channel estimation approach, compared to the conventional pilot-aided methods, for MIMO systems of different scales.

The rest of this paper is organized as follows. In Section II, we present the system model for the uplink transmission of MIMO system. In Section III, we introduce the proposed blind channel estimation framework based on VI. Then, the performance of the proposed approach is numerically evaluated in Section IV. Finally, we conclude the paper in Section V.

Notions: matrices and vectors are denoted by boldface symbols, $\mathbb{C}$ denotes the set of complex values and ${\rm N}_ +$ represents positive integer. ${\left( \cdot \right)^T}$ and ${\left( \cdot \right)^H}$ denote transpose, Hermitian respectively. The elements $i$ and $ \left( {i,j} \right)$ of the vector $a$ and matrix ${\bf{A}}$ are represented by ${a_i}$ and ${{\bf{A}}_{ij}}$, ${\bf{I }}_K$ is $K \times K$ identity matrix, and ${tr\left( \cdot \right)} $ represents the trace of matrix. ${\rho ^2}$is the transmitting power. $E\left( \cdot \right)$ and $Var\left( \cdot \right)$ denote the expectation and variance of a random variable, $\mathcal{CN}\left( {{\bf{\mu }},{\bf{\Sigma }}} \right)$ denotes circularly symmetric complex Gaussian random vectors with mean $\bf{\mu }$ and covariance matrix $\bf{\Sigma }$.

\section{SYSTEM MODEL AND PRELIMINARIES}
Consider the uplink transmission of a MIMO system under the Rayleigh fading channel, where $K$ single-antenna mobile stations (MS) are uploading data to a base station (BS) with $N$ antennas. Typical transmission operates in two phases; (i) \emph{the channel estimation phase} where known pilot sequences are transmitted to the BS to obtain the CSI, and (ii) \emph{the signal detection phase} when the estimated CSI is used to detect transmitted signals \cite{b14}. However, different from this pilot-aided approach which can result in the loss of bandwidth efficiency and pilot contamination, the proposed method in this paper aims to recover the sent signal without the help of pilots. In the channel estimation phase, instead of sending pilots, the MSs send data in an one by one order such that while one is sending during its allocated time slot, the others remain silent to maintain orthogonality. The received signal by the BS at the end of the \emph{t}th time slot can be represented as
\begin{equation}
{{\bf{y}}_t} = {\bf{H}}{{\bf{x}}_t} + {{\bf{n}}_t},
\label{eq1}
\end{equation}
where the channel matrix ${\bf{H}} \in {\mathbb{C}^{N \times K}}$ is considered to remain constant over a number of transmissions and then changes to a new state according to block fading distribution, and the  elements of $\bf{H}$ are independent random variables that follow zero-mean complex Gaussian distribution with unit variance, i.e., ${h_{ij}} \sim \mathcal{CN}(0,1)$. ${{\mathbf{x}}_t} = {\left[ {{x_1}\left( t \right),...,{x_K}\left( t \right)} \right]^T} \in {\mathbb{C}^{K \times 1}}$ is the transmitted signal vector during the \emph{t}th time slot of the channel estimation phase with only one of the elements is non-zero corresponding to the user allocated to transmit in this time slot. We also note that the independent bit streams at each MB are mapped using the same constellation map. ${{\mathbf{y}}_t} = {\left[ {{y_1}\left( t \right),...,{y_N}\left( t \right)} \right]^T} \in {\mathbb{C}^{N \times 1}}$is the received signals vector by BS, and ${{\mathbf{n}}_t}$  is the additive noises vector following zero-mean complex Gaussian distribution.

The goal of receiver design in channel estimation phase is to recover the transmitted signal ${{\mathbf{x}}_t}$ and then estimate the channel matrix $\bf{H}$ from the received signals ${{\mathbf{y}}_t}$, and then the obtained channel matrix $\bf{H}$ is utilized in the signal detection phase to detect the transmitted signals. The MAP estimation of the sent signal $\bf{x}_t$ can be expressed as
\begin{equation}
\begin{aligned}
{{{\bf{\hat x}}}_t}^{MAP} &= \mathop {\arg \max }\limits_{\bf{x}} \int\limits_{\bf{H}} {p\left( {{{\bf{x}}_t},{\bf{H}}|{{\bf{y}}_t}} \right){\rm{ }}} d{\bf{H}}\\
{\rm{        }} &= \mathop {\arg \max }\limits_{\bf{x}} \int\limits_{\bf{H}} {p\left( {{{\bf{x}}_t}|{\bf{H}},{{\bf{y}}_t}} \right)} p\left( {{\bf{H}}|{{\bf{y}}_t}} \right)d{\bf{H}},
\end{aligned}
\label{eq2}
\end{equation}
where the second equation can be derived by applying Bayesian criterion. Unfortunately, due to the high dimensionality of channel matrix $\bf{H}$, it is infeasible to directly calculate ${{{\bf{\hat x}}}_t}^{MAP}$ by \eqref{eq2}. Hence, we propose an approximate inference method instead that approximates ${{{\bf{\hat x}}}_t}^{MAP}$ and estimates $\bf{H}$  simultaneously via a neural network based variance inference framework which will be introduced in details in next section. 

\section{OPTIMIZATION FRAMEWORK FOR BLIND ESTIMATION}

In this section, we present the proposed method for simultaneously estimating the transmitted symbols and channel matrix without the assistance of pilots. Note that the log-likelihood function of the received signal vector ${{\mathbf{y}}_t}$ during the $t$th time slot of the channel estimation phase can be written as 

\begin{subequations}  \label{eq3}
\begin{align} 
\log p\left( {{{\bf{y}}_t}} \right) &=  \int_{{{\bf{x}}_t},{\bf{H}}} \log p\left( {{{\bf{y}}_t}} \right) \cdot {q\left( {{{\bf{x}}_t},{\bf{H}}|{{\bf{y}}_t}} \right)}d{{\bf{x}}_t}d{\bf{H}} \\ 
&\nonumber =\int_{{{\bf{x}}_t},{\bf{H}}} {q\left( {{{\bf{x}}_t},{\bf{H}}|{{\bf{y}}_t}} \right)}  \\&\cdot \log \frac{{p\left( {{{\bf{x}}_t},{\bf{H}},{{\bf{y}}_t}} \right)}}{{q\left( {{{\bf{x}}_t},{\bf{H}}|{{\bf{y}}_t}} \right)}}\frac{{q\left( {{{\bf{x}}_t},{\bf{H}}|{{\bf{y}}_t}} \right)}}{{p\left( {{{\bf{x}}_t},{\bf{H}}|{{\bf{y}}_t}} \right)}}d{{\bf{x}}_t}d{\bf{H}}\\ 
 &\nonumber = {E_{{{\bf{x}}_t}{\bf{,H}}\sim q\left( {{{\bf{x}}_t},{\bf{H}}|{{\bf{y}}_t}} \right)}}\left[ {\log \frac{{p\left( {{{\bf{x}}_t},{\bf{H}},{{\bf{y}}_t}} \right)}}{{q\left( {{{\bf{x}}_t},{\bf{H}}|{{\bf{y}}_t}} \right)}}} \right] \\&+ \underbrace {{E_{{{\bf{x}}_t}{\bf{,H}}\sim q\left( {{{\bf{x}}_t},{\bf{H}}|{{\bf{y}}_t}} \right)}}\left[ {\log \frac{{q\left( {{{\bf{x}}_t},{\bf{H}}|{{\bf{y}}_t}} \right)}}{{p\left( {{{\bf{x}}_t},{\bf{H}}|{{\bf{y}}_t}} \right)}}} \right]}_{Kullback - Leibler \enspace divergence}  
\end{align}
\end{subequations}
where an auxiliary distribution $q\left( {{{\bf{x}}_t},{\bf{H}}|{{\bf{y}}_t}} \right)$ is introduced in (3a), and (3b) is derived by expanding ${p\left( {{{\bf{y}}_t}} \right)}$. We note that the \emph{Kullback-Leibler divergence} (KL-D)\cite{b16} between the posterior distribution $p\left({{{\bf{x}}_t},{\bf{H}}|{{\bf{y}}_t}} \right)$ and the auxiliary distribution $q\left( {{{\bf{x}}_t},{\bf{H}}|{{\bf{y}}_t}} \right)$ is non-negative, therefore a lower bound of the intractable likelihood function of received signal ${{\mathbf{y}}_t}$ is derived through VI framework. We have
\begin{equation}
\begin{aligned}
\log p\left( {{{\bf{y}}_t}} \right) &\ge {E_{{{\bf{x}}_t}{\bf{,H}}\sim q\left( {{{\bf{x}}_t},{\bf{H}}|{{\bf{y}}_t}} \right)}}\left[ {\log \frac{{p\left( {{{\bf{y}}_t},{\bf{H}},{{\bf{x}}_t}} \right)}}{{q\left( {{{\bf{x}}_t},{\bf{H}}|{{\bf{y}}_t}} \right)}}} \right] \\& \buildrel \Delta \over =  - L\left( q \right).
\end{aligned}
\label{eq4}
\end{equation}

\begin{figure}
	\includegraphics[width=3.5in]{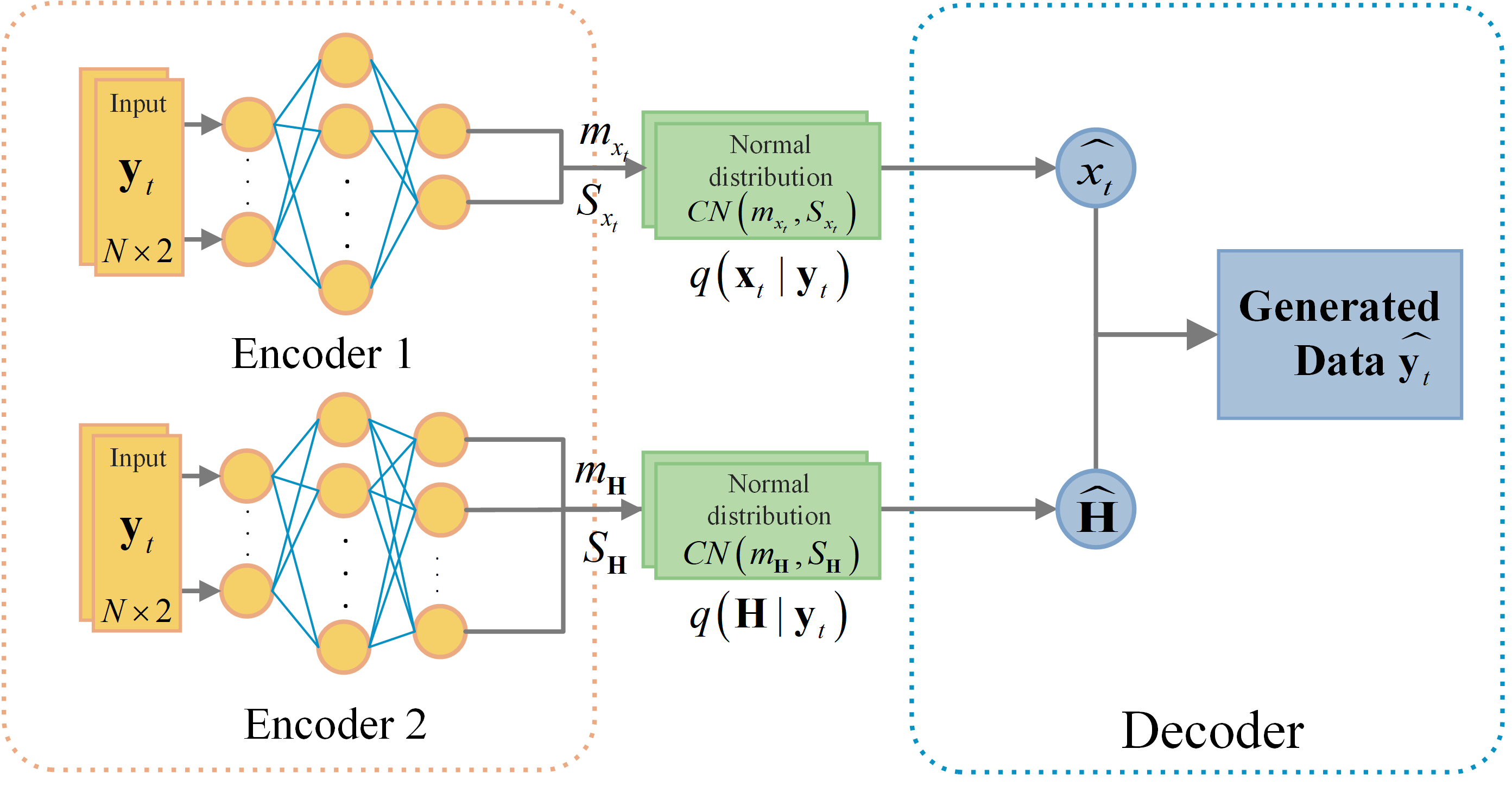}
	\caption{The blind estimation framework structure. }
	\label{fig5}
\end{figure}

Since $\log p\left( {{{\bf{y}}_t}} \right)$ is a unknown constant by (3c), we can minimize the  KL-divergence ${D_{KL}}\left( {p\left( {{{\bf{x}}_t},{\bf{H}}|{{\bf{y}}_t}} \right)||q\left( {{{\bf{x}}_t},{\bf{H}}|{{\bf{y}}_t}} \right)} \right)$ by maximizing  the lower bound $- L\left( q \right)$ over the parameters ${{\bf{x}}_t}$ and ${\bf{H}}$, by which the distribution $q\left( {{{\bf{x}}_t},{\bf{H}}|{{\bf{y}}_t}} \right)$ approximates the posterior distribution $p\left( {{{\bf{x}}_t},{\bf{H}}|{{\bf{y}}_t}} \right)$. Thus we can obtain the maximum likelihood estimation of channel matrix ${\bf{H}}$ and transmitted signals ${{\bf{x}}_t}$ with $q\left( {{{\bf{x}}_t},{\bf{H}}|{{\bf{y}}_t}} \right)$. We exploit the mean-field approximation\cite{b17} to further simplify the maximum likelihood estimation problem by assuming $q\left( {{{\bf{x}}_t},{\bf{H}}|{{\bf{y}}_t}} \right) = q\left( {{{\bf{x}}_t}|{{\bf{y}}_t}} \right)q\left( {{\bf{H}}|{{\bf{y}}_t}} \right)$. We also assume $q\left( {{\bf{H}}|{{\bf{y}}_t}} \right)$ and $q\left( {{{\bf{x}}_t}|{{\bf{y}}_t}} \right)$  follow complex Gaussian distribution as follows
\begin{equation}
\begin{aligned}
q\left( {{{\bf{x}}_t}|{{\bf{y}}_t}} \right)\sim \mathcal{CN}\left( {{m_{{{\bf{x}}_t}}},{S_{{{\bf{x}}_t}}}} \right),\\q\left( {{\bf{H}}|{{\bf{y}}_t}} \right)\sim \mathcal{CN}\left( {{m_{{{\bf{H}}}}},{S_{{{\bf{H}}}}}} \right),
\end{aligned}
\label{eq5}
\end{equation}
where ${m_{{{\bf{x}}_t}}}$, ${m_{{{\bf{H}}}}}$, $S_{{{\bf{x}}_t}}$, $S_{{{\bf{H}}}}$ denote  the means and variances of the Gaussian distribution respectively. We obtain these parameters by two trainable neural networks denoted by  \emph{g} and  \emph{f}  respectively as shown in Fig.~\ref{fig5}
\begin{equation}
\begin{aligned}
\left[ {{m_{{{\bf{x}}_t}}},{S_{{{\bf{x}}_t}}}} \right]{\rm{ = }}g\left( {{{\bf{y}}_t}\left| \phi  \right.} \right),\\{\rm{ }}\left[ {{m_{{{\bf{H}}}}},{S_{{{\bf{H}}}}}} \right] = f\left( {{{\bf{y}}_t}\left| \varphi  \right.} \right),
\end{aligned}
\label{eq6}
\end{equation}
referred to as $Encoder1$ and $Encoder2$ of the proposed blind channel estimation framework. The inputs to these two encoders are the received signals ${{\bf{y}}_t}$, where the two channels corresponding to the real and imaginary elements. Both $Encoder1$ and $Encoder2$ have two fully connected layer with $\phi$ and $\varphi$ being the trainable parameters of the networks. The estimated ${m_{{{\bf{x}}_t}}}$, ${m_{{{\bf{H}}}}}$, $S_{{{\bf{x}}_t}}$, $S_{{{\bf{H}}}}$ are used to generate the complex Gaussian distribution $\mathcal{CN}\left( {{m_{{{\bf{x}}_t}}},{S_{{{\bf{x}}_t}}}} \right)$ and $\mathcal{CN}\left( {{m_{{{\bf{H}}_t}}},{S_{{{\bf{H}}_t}}}} \right)$ respectively, from which we sample the estimation of ${\bf{x}}_k$ and ${\bf{H}}$, denoted by $\widehat {{{\bf{x}}_t}}$, ${\widehat{{\bf{H}}}}$. Finally, we generate $\widehat {{{\bf{y}}_t}}$ by multiplying $\widehat {{{\bf{x}}_t}}$ and ${\widehat{{\bf{H}}}}$, which is referred to as $Decoder$ of the proposed framework. We train the whole network structure by minimizing the objective function $L\left( q \right)$ where  $q\left( {{{\bf{x}}_t},{\bf{H}}|{{\bf{y}}_t}} \right) = q\left( {{{\bf{x}}_t}|{{\bf{y}}_t}} \right)q\left( {{\bf{H}}|{{\bf{y}}_t}} \right)$, that is the product of the two normal distribution generate by $Encoder1$ and $Encoder2$, $L\left( q \right)$ can be written as
\begin{equation}
\begin{aligned}
L\left( q \right) &= \int_{{{\bf{x}}_t},{\bf{H}}\sim q\left( {{{\bf{x}}_t},{\bf{H}}|{{\bf{y}}_t}} \right)} {q\left( {{{\bf{x}}_t},{\bf{H}}|{{\bf{y}}_t}} \right)} \log \frac{{q\left( {{{\bf{x}}_t},{\bf{H}}|{{\bf{y}}_t}} \right)}}{{p\left( {{{\bf{x}}_t},{\bf{H}},{{\bf{y}}_t}} \right)}}d{{\bf{x}}_t}d{\bf{H}}\\
 &= \underbrace {{E_{{\bf{H}},{{\bf{x}}_t}\sim q\left( {{{\bf{x}}_t},{\bf{H}}|{{\bf{y}}_t}} \right)}}\left[ {\log \frac{{q\left( {{{\bf{x}}_t}{\bf{|}}{{\bf{y}}_t}} \right)}}{{p\left( {{{\bf{x}}_t}} \right)}}} \right]}_{los{s_1}} \\&+ \underbrace {{E_{{\bf{H}},{{\bf{x}}_t}\sim q\left( {{{\bf{x}}_t},{\bf{H}}|{{\bf{y}}_t}} \right)}}\left[ {\log \frac{{q\left( {{\bf{H|}}{{\bf{y}}_t}} \right)}}{{p\left( {\bf{H}} \right)}}} \right]}_{los{s_2}} \\&- \underbrace {{E_{{\bf{H}},{{\bf{x}}_t}\sim q\left( {{{\bf{x}}_t},{\bf{H}}|{{\bf{y}}_t}} \right)}}\left[ {\log p\left( {{{\bf{y}}_t}{\bf{|H,}}{{\bf{x}}_t}} \right)} \right]}_{los{s_3}}.
\end{aligned}
\label{eq7}
\end{equation}

We have
\begin{equation}
\begin{aligned}
&los{s_1} \\&= {E_{{{\bf{x}}_t}\sim q\left( {{{\bf{x}}_t}|{{\bf{y}}_t}} \right)}}\left[ {\log \frac{{q\left( {{{\bf{x}}_t}|{{\bf{y}}_t}} \right)}}{{p\left( {{{\bf{x}}_t}} \right)}}} \right]\\&={E_{{{\bf{x}}_t}\sim q\left( {{{\bf{x}}_t}|{{\bf{y}}_t}} \right)}}\left[ {\log q\left( {{{\bf{x}}_t}|{{\bf{y}}_t}} \right)} \right]-{E_{{{\bf{x}}_t}\sim q\left( {{{\bf{x}}_t}|{{\bf{y}}_t}} \right)}}\left[ {\log p\left( {{{\bf{x}}_k}} \right)} \right].
\end{aligned}
\label{eq8}
\end{equation}

Note that the term ${E_{{{\bf{x}}_t}\sim q\left( {{{\bf{x}}_t}|{{\bf{y}}_t}} \right)}}\left[ {\log {q\left( {{{\bf{x}}_t}|{{\bf{y}}_t}} \right)}} \right]$ is the entropy of multivariate normal distribution ${q\left( {{{\bf{x}}_t}|{{\bf{y}}_t}} \right)}\sim \mathcal{CN}\left( {{m_{{{\bf{x}}_t}}},{S_{{{\bf{x}}_t}}}} \right)$. We have
\begin{equation}
\begin{aligned}
{E_{{{\bf{x}}_t}\sim q\left( {{{\bf{x}}_t}|{{\bf{y}}_t}} \right)}}\left[ {\log q\left( {{{\bf{x}}_t}|{{\bf{y}}_t}} \right)} \right] =  - \frac{1}{2}\log \left| {{S_{{{\bf{x}}_t}}}} \right|+C_1,
\end{aligned}
\label{eq9}
\end{equation}
where $C$ is a constant term and we use $C_i$, $i \in {{\rm N}_ + }$, to denote different constants in the sequel. And the term ${E_{{{\bf{x}}_t}\sim q\left( {{{\bf{x}}_t}|{{\bf{y}}_t}} \right)}}\left[ {\log {p\left( {{{\bf{x}}_t}} \right)}} \right]$ in \eqref{eq8} can be written as 
\begin{equation}
\begin{aligned}
 - {E_{{{\bf{x}}_t}\sim q\left( {{{\bf{x}}_t}|{{\bf{y}}_t}} \right)}}\left[ {\log {p\left( {{{\bf{x}}_t}} \right)}} \right] & = \frac{1}{{2{\rho ^2}}}{E_{{{\bf{x}}_t}\sim q\left( {{{\bf{x}}_t}|{{\bf{y}}_t}} \right)}}\left[ {{{\bf{x}}_t}^H{{\bf{x}}_t}} \right]+C_2 \\&= \frac{1}{{2{\rho ^2}}}\left( {tr({S_{{{\bf{x}}_t}}}) + {m_{{{\bf{x}}_t}}}^T{m_{{{\bf{x}}_t}}}} \right)+C_2,
\end{aligned}
\label{eq10}
\end{equation}
where the discrete sample space constrain of ${{\bf{x}}_t}$ is relaxed to be continuous via the assumption $p\left( {{{\bf{x}}_t}} \right)\sim \mathcal{CN}\left( {0,2{\rho ^2}{\bf{I }}_K} \right)$ for the convenience of the calculation. And hence the $los{s_1}$ can be rewritten as
\begin{equation}
\begin{aligned}
los{s_1}=\frac{1}{{2{\rho ^2}}}\left( {tr({S_{{{\bf{x}}_t}}}) + {m_{{{\bf{x}}_t}}}^T{m_{{{\bf{x}}_t}}}} \right) - \frac{1}{2}\log \left| {{S_{{{\bf{x}}_t}}}} \right|+C_3.
\end{aligned}
\label{eq11}
\end{equation}

The detailed proof of equations \eqref{eq9}-\eqref{eq11} can be found in Appendix I. Similarly, we have
\begin{equation}
\begin{aligned}
&los{s_2} \\&= {E_{{\bf{H}}\sim q\left( {{\bf{H}}|{{\bf{y}}_t}} \right)}}\left[ {\log \frac{{q\left( {{\bf{H}}|{{\bf{y}}_t}} \right)}}{{p\left( {\bf{H}} \right)}}} \right]
\\&={E_{{\bf{H}}\sim q\left( {{\bf{H}}|{{\bf{y}}_k}} \right)}}\left[ {\log q\left( {{\bf{H}}|{{\bf{y}}_k}} \right)} \right]  - {E_{{\bf{H}}\sim q\left( {{\bf{H}}|{{\bf{y}}_k}} \right)}}\left[ {\log p\left( {\bf{H}} \right)} \right].
\end{aligned}
\label{eq12}
\end{equation}

Following the same procedure of deriving \eqref{eq9}-\eqref{eq11}, we have
\begin{equation}
\begin{aligned}
 {E_{{\bf{H}}\sim q\left( {{\bf{H}}|{{\bf{y}}_k}} \right)}}\left[ {\log q\left( {{\bf{H}}|{{\bf{y}}_k}} \right)} \right] =  - \frac{1}{2}\log \left| {{S_{\bf{H}}}} \right|+C_4,
\end{aligned}
\label{eq13}
\end{equation}
and the term ${E_{{{\bf{H}}}\sim q\left( {{{\bf{H}}}|{{\bf{y}}_t}} \right)}}\left[ {\log {p\left( {{{\bf{H}}}} \right)}} \right]$ in \eqref{eq12} can be written as 
\begin{equation}
\begin{aligned}
 - {E_{{\bf{H}}\sim q\left( {{\bf{H}}|{{\bf{y}}_k}} \right)}}\left[ {\log p\left( {\bf{H}} \right)} \right] &= \frac{1}{2}{E_{{\bf{H}}\sim q\left( {{\bf{H}}|{{\bf{y}}_t}} \right)}}\left[ {{{\bf{H}}^H}{\bf{H}}} \right]+C_5\\&= \frac{1}{2} tr({S_{\bf{H}}}) + \frac{1}{2}{m_{\bf{H}}}^T{m_{\bf{H}}}+C_5.
\end{aligned}
\label{eq14}
\end{equation}
Hence we rewritten $los{s_2}$  as
\begin{equation}
\begin{aligned}
los{s_2} =  \frac{1}{2}_tr({S_{\bf{H}}}) +  \frac{1}{2}{m_{\bf{H}}}^T{m_{\bf{H}}} - \frac{1}{2}\log \left| {{S_{\bf{H}}}} \right|+C_6.
\end{aligned}
\label{eq15}
\end{equation}

We employ Monte Carlo method to compute the $los{s_3}$ in  \eqref{eq7}, we have
\begin{equation}
\begin{aligned}
los{s_3}&=- {E_{{\bf{H}}\sim q\left( {{\bf{H}}|{{\bf{y}}_t}} \right),{\bf{x}}\sim q\left( {{{\bf{x}}_t}|{{\bf{y}}_t}} \right)}}\left[ {\log p\left( {{{\bf{y}}_t}|{\bf{H}},{{\bf{x}}_t}} \right)} \right]\\& \approx \frac{1}{L}\sum\limits_{l = 1}^L {}  {tr({\widehat{{\bf{H}}}_{\rm{l}}}{S_{{{\bf{x}}_t}}}{\widehat{{\bf{H}}}_{\rm{l}}}^H)} \\& + \frac{1}{L}\sum\limits_{l = 1}^L {}  {{{({\widehat{{\bf{H}}}_{\rm{l}}}{m_{{{\bf{x}}_t}}} - {{\bf{y}}_t})}^H}({\widehat{{\bf{H}}}_{\rm{l}}}{m_{{{\bf{x}}_t}}} - {{\bf{y}}_t})},
\end{aligned}
\label{eq17}
\end{equation}
where ${\widehat{{\bf{H}}}_{\rm{l}}}$ are sampled from ${q\left( {{\bf{H}}|{{\bf{y}}_t}} \right)}\sim \mathcal{CN}\left( {{m_{{{\bf{H}}_t}}},{S_{{{\bf{H}}_t}}}} \right)$ and $L$ is the number of sample points. The proof of \eqref{eq17} can also be found in Appendix.

We note that \eqref{eq8} and \eqref{eq12} represent the KL-D between the posterior distributions generated by the proposed neural network and the actual priors distribution of ${{\bf{x}}_t}$ and ${{\bf{H}}}$ respectively. $los{s_3}$ represents the reconstruction error  $\widehat {{{\bf{y}}_t}}$ with the variational distributions ${q\left( {{{\bf{x}}_t}|{{\bf{y}}_t}} \right)}$ and ${q\left( {{\bf{H}}|{{\bf{y}}_t}} \right)}$. Hence minimizing the objective function $L\left( q \right)=los{s_1}+los{s_2}+los{s_3}$ pushes the generated posterior distribution approach the prior distribution and the reconstructed signal $\widehat {{{\bf{y}}_t}}$ to approach the actual received signal ${{{\bf{y}}_t}}$. Thus when the loss function converges, reasonable estimation results about ${\bf{H}}$ and decision results ${{\bf{x}}_t}$ about can be obtained.

\section{NUMERICAL ANALYSIS}

\begin{figure}{}
\centering
\subfigure[signals before equalization with QPSK]{\includegraphics[width=3.5cm]{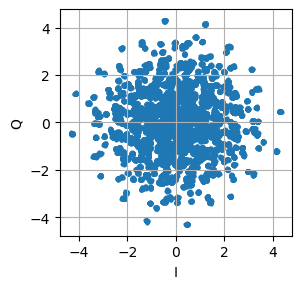}} 
\subfigure[signals after equalization with QPSK]{\includegraphics[width=3.7cm]{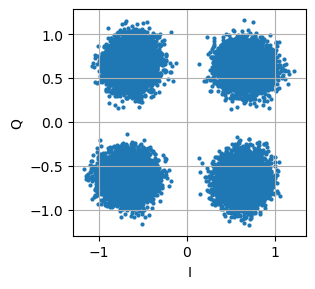}}
\\ 
\centering
\subfigure[signals before equalization with 16QAM]{\includegraphics[width=3.6cm]{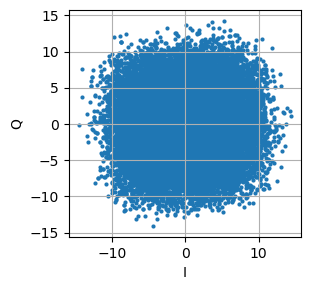}}
\subfigure[signals after equalization with 16QAM]{\includegraphics[width=3.5cm]{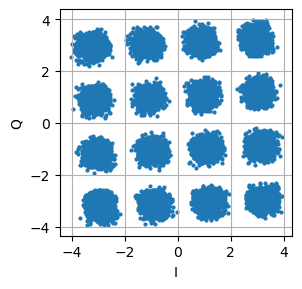}}
\caption{Blind equalization result of receiver setup with 4 antennas in medium SNR scenario: (a) QPSK modulation before blind equalization; (b) QPSK modulation after blind equalization; (c) 16QAM modulation before blind equalization; (d) 16QAM modulation after blind equalization.} 
\label{fig1}
\end{figure}
In this section, we numerically evaluate the performance of our proposed blind channel estimation framework. A MIMO system in which $K = 4$ users are communicating with a BS through QPSK/16QAM modulation is considered here. Both $Encoder1$ and $Encoder2$ of the proposed framework are fully connected neural networks, each of which consists of an input layer, a 16 node hidden layer with tanh activation and an output layer with tanh activation. The Adam optimizer\cite{b18} with an initial learning rate of 0.05 is used to train the whole network. The pilot-aided channel estimation methods \cite{b7} is used as the benchmark.

We first demonstrate in Fig.~\ref{fig1} the equalization results by the proposed blind estimation method through the constellation graph. It is seen from Fig.~\ref{fig1}(a) that the transmitted signals during the signal detection phase interfered by multi-path fading are overlapped with each other, and the proposed method is able to separate the overlapped signal points as shown in Fig.~\ref{fig1}(b) which implies the effectiveness of our blind estimation algorithm. When the modulation order increased from QPSK to 16QAM, our algorithm still works effectively as shown in Fig.~\ref{fig1}(c) and Fig.~\ref{fig1}(d), which demonstrates the generalization ability of our method among different order modulation schemes.
\begin{figure}
	\includegraphics[width=3.4in]{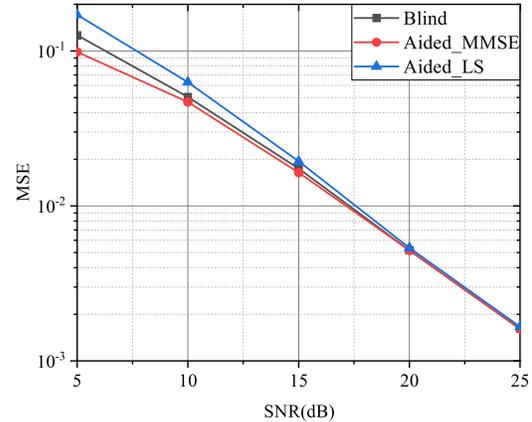}
	\caption{The MSE with respect to SNR for $K = 4$ users and $N = 4$ receive antennas with QPSK modulation.}
	\label{fig2}
\end{figure}
\begin{figure}
	\includegraphics[width=3.4in]{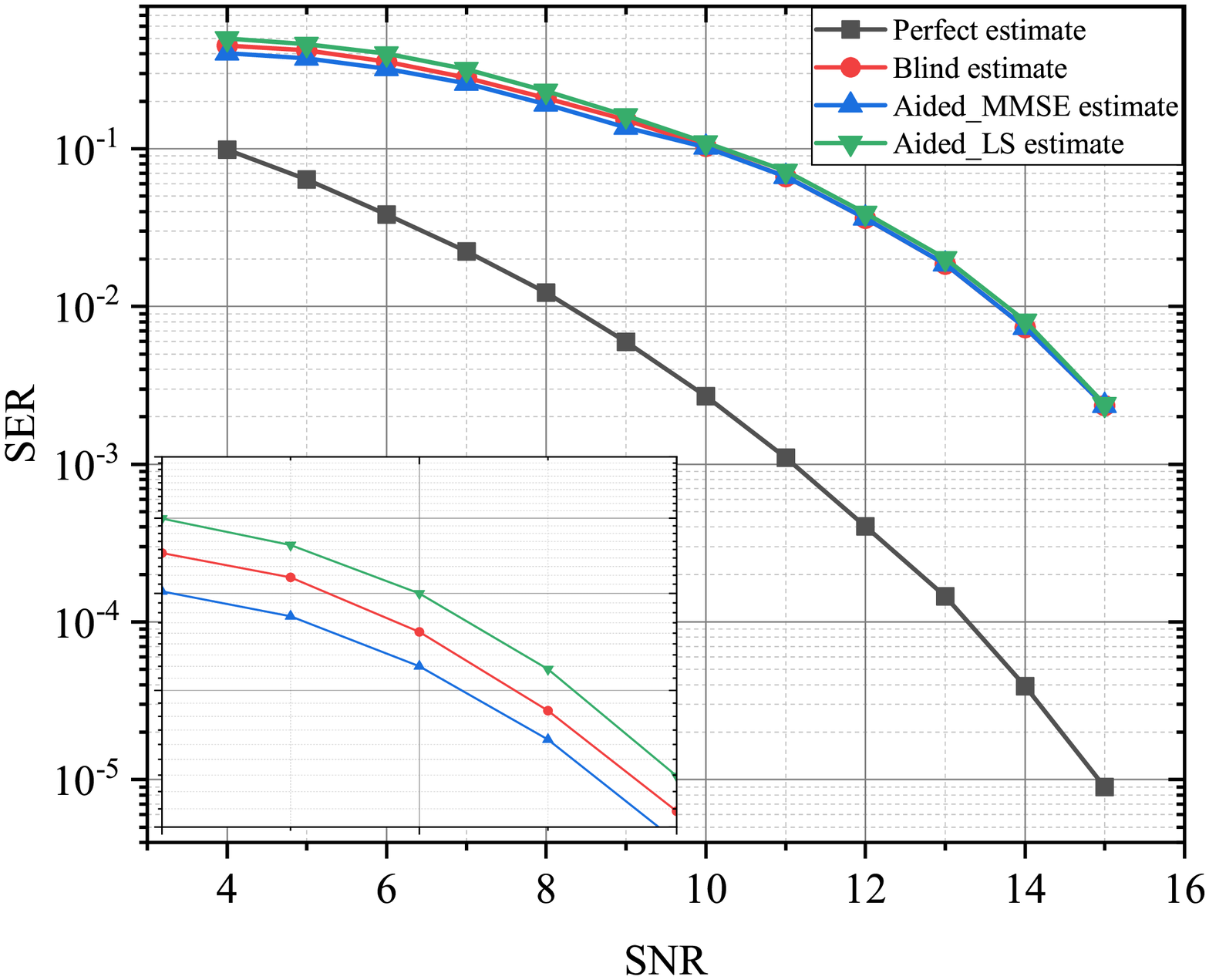}
	\caption{The SER performance with respect to SNR for $K = 4$ users and $N = 4$  antennas with QPSK modulation. }
	\label{fig3}
\end{figure}
\begin{figure}
	\includegraphics[width=3.4in]{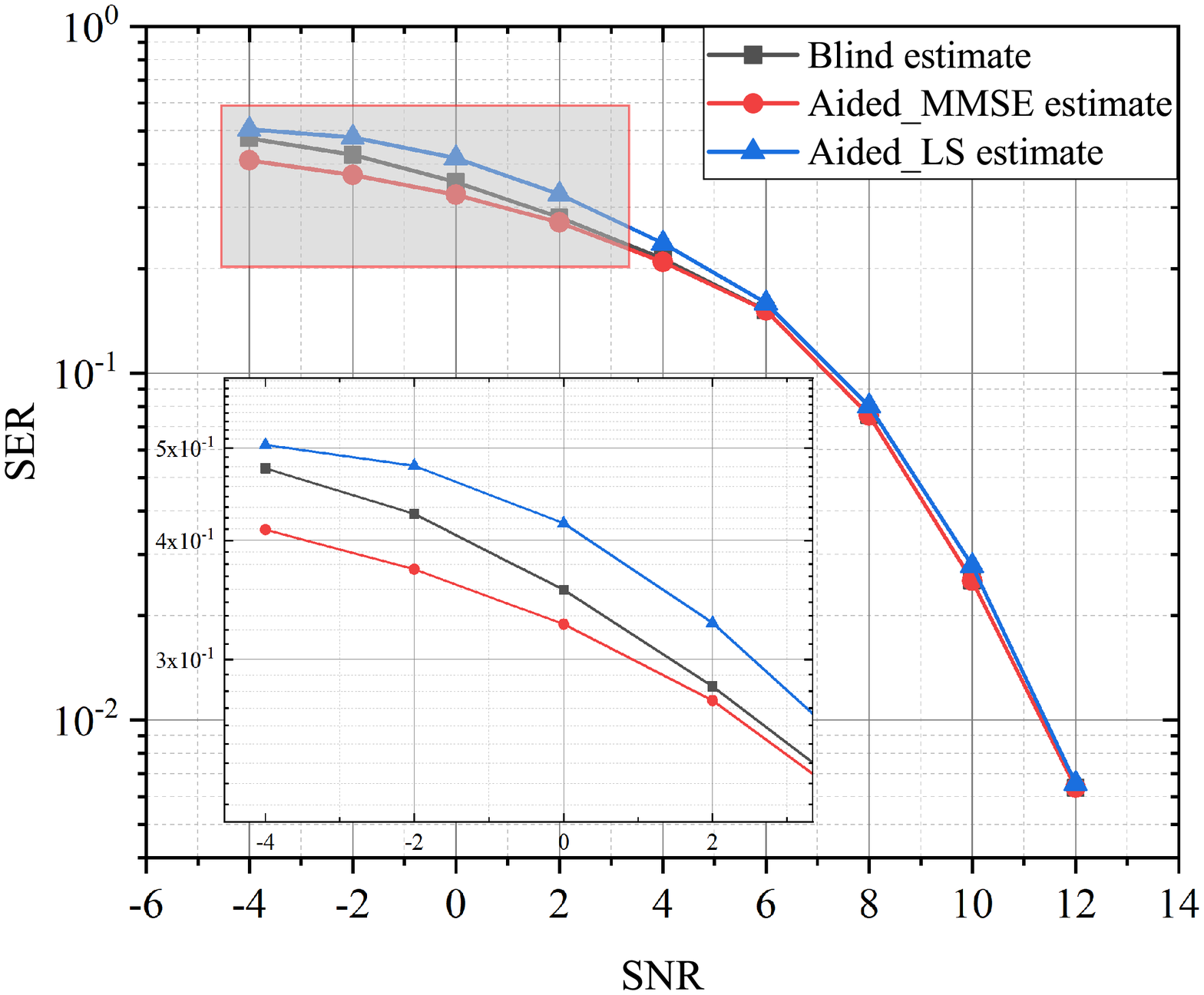}
	\caption{The SER performance with respect to SNR for $K = 4$ users and $N = 40$  antennas with QPSK modulation. }
	\label{fig4}
\end{figure}

Then we compare the channel estimation performance of conventional pilot-aided approach with minimum mean squared error algorithm (Aided-MMSE), pilot-aided with least square algorithm (Aided-LS), and the proposed method in terms of through mean square error (MSE) depicted in Fig.~\ref{fig2}. We can observe that in the low-to-medium SNR region, knowing the exact pilot symbols and utilizing the statistics of channel to eliminate the AWGN, Aided-MMSE achieves the best performance, while the estimation error of our blind method is slightly higher due to the strong AWGN interference. However, the proposed method still outperforms Aided-LS which does not use the channel statistics even without the assistance of pilots. It can also be observed that in the medium-to-high SNR region, the estimation performance of the three methods is very close which validates the effectiveness of the proposed channel estimation method.

We also simulate the symbol error rate (SER) in signal detection phase, where the maximum-likelihood detector (MLD) with CSI estimated in channel estimation phase is employed, we also use the results by the MLD with perfect CSI estimation as a benchmark. As shown in Fig.~\ref{fig3}, it can be seen that the signal detection performance of proposed blind method in terms of SER approaches the Aided-MMSE detector while slightly outperforms the Aided-LS detector in low-to-medium SNR scenario due to the channel estimation performance as shown in Fig.~\ref{fig2}.  With the increase of SNR, the performance gap between the three methods becomes neglectable. Meanwhile, the performance gap between the detector with perfect CSI and the three detectors with estimated CSI decreases as well due to the increase of estimation accuracy. We emphasize that our method could save spectrum resources and mitigate the pilot contamination since there is no need for pilot signals. We also evaluate signal detection performance when there are more receiver antennas ($K = 40$) in Fig.~\ref{fig4} and observe the same performance trend, which proves the compatibility of our blind method with MIMO systems of different scales.

\section{CONCLUSIONS}
In this paper, we proposed a novel blind channel estimation approach for MIMO systems experiencing Rayleigh fading by exploiting variational inference and neural network. We derived a lower bound to the intractable log-likelihood of received signal by introducing an auxiliary which is generated by a neural network based framework. By training the neural network to maximize the lower bound, the auxiliary posterior distribution closely approaches the real distribution, by sampling from which the estimation of CSI can be obtained. We then numerically compared the proposed blind estimation method with the conventional pilot-aided methods in terms of the channel estimation error and SER of the detected signals which demonstrated that the proposed method outperforms the pilot-aided scheme with LS algorithm, and closely approaches the performance of the pilot-aided scheme with MMSE algorithm while saving spectrum resource and mitigating the pilot contamination problem.

\section*{APPENDIX I}
In this section, the detailed derivation of each term in the blind estimation framework \eqref{eq7} is given.
For ${E_{{{\bf{x}}_t}\sim q\left( {{{\bf{x}}_t}|{{\bf{y}}_t}} \right)}}\left[ {\log q\left( {{{\bf{x}}_t}|{{\bf{y}}_t}} \right)} \right]$, the probability density function (PDF) of the multivariate normal distribution $q\left( {{\bf{H}}|{{\bf{y}}_t}} \right)\sim \mathcal{CN}\left( {{m_{{{\bf{H}}_t}}},{S_{{{\bf{H}}_t}}}} \right)$ can be written as
\begin{equation}
\begin{aligned}
q\left( {{\bf{H}}|{{\bf{y}}_t}} \right) &= {\left( {2\pi } \right)^{{\rm{ - }}\frac{N \times K}{2}}}{\left| {{S_{\bf{H}}}} \right|^{{\rm{ - }}\frac{1}{2}}} \\&\times \exp \left[ {{\rm{ - }}\frac{1}{2}{{\left( {{\bf{H}} - {m_{\bf{H}}}} \right)}^H}{S_{\bf{H}}}^{{\rm{ - 1}}}\left( {{\bf{H}} - {m_{\bf{H}}}} \right)} \right]d{{\bf{x}}_t}d{\bf{H}},
\end{aligned}
\label{eq18}
\end{equation}
and substituting the PDF \eqref{eq18} into the entropy of normal distribution  ${E_{{{\bf{x}}_t}\sim q\left( {{{\bf{x}}_t}|{{\bf{y}}_t}} \right)}}\left[ {\log q\left( {{{\bf{x}}_t}|{{\bf{y}}_t}} \right)} \right]$, we have
\begin{equation}
\begin{aligned}
&\int_{x,{\bf{H}}} {q\left( {{\bf{H}}|{{\bf{y}}_t}} \right)} \log q\left( {{\bf{H}}|{{\bf{y}}_t}} \right)d{{\bf{x}}_t}d{\bf{H}}\\
 &= \int_{x,{\bf{H}}} {q\left( {{\bf{H}}|{{\bf{y}}_t}} \right)} \log {\left( {2\pi } \right)^{{\rm{ - }}\frac{N \times K}{2}}}{\left| {{S_{\bf{H}}}} \right|^{{\rm{ - }}\frac{1}{2}}} \\&\times \exp \left[ {{\rm{ - }}\frac{1}{2}{{\left( {{\bf{H}} - {m_{\bf{H}}}} \right)}^H}{S_{\bf{H}}}^{{\rm{ - 1}}}\left( {{\bf{H}} - {m_{\bf{H}}}} \right)} \right]d{{\bf{x}}_t}d{\bf{H}}\\
&{\rm{ = }} - \frac{N \times K}{2}\left( {\log 2\pi  + 1} \right) - \frac{1}{2}\log \left| {{S_{\bf{H}}}} \right|\\&=  - \frac{1}{2}\log \left| {{S_{\bf{H}}}} \right|+C_4,
\end{aligned}
\label{eq19}
\end{equation}
where the constant term $- \frac{N \times K}{2}\left( {\log 2\pi  + 1} \right)$ is a constant and denoted by $C_4$, and  ${E_{{\bf{H}}\sim q\left( {{\bf{H}}|{{\bf{y}}_t}} \right)}}\left[ {q\left( {{\bf{H}}|{{\bf{y}}_t}} \right)} \right]$ can be derived directly in the same way
\begin{equation}
\begin{aligned}
\int_{{{\bf{x}}_t},{\bf{H}}} {q\left( {{{\bf{x}}_t}|{{\bf{y}}_t}} \right)} \log q\left( {{{\bf{x}}_t}|{{\bf{y}}_t}} \right)d{{\bf{x}}_t}d{\bf{H}}  =  - \frac{1}{2}\log \left| {{S_{{{\bf{x}}_t}}}} \right|+C_1.
\end{aligned}
\label{eq19}
\end{equation}

For $-{E_{{{\bf{x}}_t}\sim q\left( {{{\bf{x}}_t}|{{\bf{y}}_t}} \right)}}\left[ {\log p\left( {{{\bf{x}}_t}} \right)} \right]$, the PDF of $p\left( {{{\bf{x}}_t}} \right)\sim \mathcal{CN}\left( {0,2{\rho ^2}{{\bf{I }}_K}} \right)$ can be written as 
\begin{equation}
\begin{aligned}
p\left( {{{\bf{x}}_t}} \right) &= {\left( {2\pi } \right)^{{\rm{ - }}\frac{P}{2}}}{\left| {2{\rho ^2}{{\bf{I }}_K}} \right|^{{\rm{ - }}\frac{1}{2}}}\\&\times\exp \left[ {{\rm{ - }}{{\bf{x}}_t}^H{{\left( {2{\rho ^2}{{\bf{I }}_K}} \right)}^{{\rm{ - 1}}}}{{\bf{x}}_t}} \right]d{{\bf{x}}_t}d{\bf{H}},
\end{aligned}
\label{eq20}
\end{equation}
and substituting the PDF \eqref{eq20} into $-{E_{{{\bf{x}}_t}}}\left[ {\log p\left( {{{\bf{x}}_t}} \right)} \right]$, we have
\begin{equation}
\begin{aligned}
 &- \int_{{{\bf{x}}_t},{\bf{H}}} {q\left( {{{\bf{x}}_t}|{{\bf{y}}_t}} \right)} \log p\left( {{{\bf{x}}_t}} \right)d{{\bf{x}}_t}d{\bf{H}}\\&{\rm{ = }} - \int_{{{\bf{x}}_t},{\bf{H}}} {q\left( {{{\bf{x}}_t}|{{\bf{y}}_t}} \right)} \log {\left( {2\pi } \right)^{{\rm{ - }}\frac{K}{2}}}{\left| {2{\rho ^2}{{\bf{I }}_K}} \right|^{{\rm{ - }}\frac{1}{2}}}\\&\times\exp \left[ {{\rm{ - }}{{\bf{x}}_t}^H{{\left( {2{\rho ^2}{{\bf{I }}_K}} \right)}^{{\rm{ - 1}}}}{{\bf{x}}_t}} \right]d{{\bf{x}}_t}d{\bf{H}}\\&
 = \int_{{{\bf{x}}_t},{\bf{H}}} {q\left( {{{\bf{x}}_t}|{{\bf{y}}_t}} \right)} \left( {\log {{\left( {2\pi } \right)}^{\frac{K}{2}}}{{\left| {2{\rho ^2}{{\bf{I }}_K}} \right|}^{\frac{1}{2}}} + \frac{1}{{2{\rho ^2}}}{{\bf{x}}_t}^H{{\bf{x}}_t}} \right)d{{\bf{x}}_t}d{\bf{H}}\\&
  = \frac{1}{{2{\rho ^2}}}{E_{{{\bf{x}}_t}\sim q\left( {{{\bf{x}}_t}|{{\bf{y}}_t}} \right)}}\left[ {{{\bf{x}}_t}^H{{\bf{x}}_t}} \right]+C_2,
\end{aligned}
\label{eq21}
\end{equation}
where the expectation of ${\log {{\left( {2\pi } \right)}^{\frac{N}{2}}}{{\left| {2{\rho ^2}{{\bf{I }}_K}} \right|}^{\frac{1}{2}}}}$ is constant and detonated by $C_2$, and since $E\left[ {{x^2}} \right] = Var\left( x \right) + {E^2}\left[ x \right]$, \eqref{eq21} can be represented as
\begin{equation}
\begin{aligned}
\frac{1}{{2{\rho ^2}}}{E_{{{\bf{x}}_t}\sim q\left( {{{\bf{x}}_t}|{{\bf{y}}_t}} \right)}}\left[ {{{\bf{x}}_t}^H{{\bf{x}}_t}} \right]&{\rm{                                        }} = \frac{1}{{2{\rho ^2}}}\left( {Var({{\bf{x}}_t}) + {E^2}\left[ {{{\bf{x}}_t}} \right]} \right)\\&
{\rm{                                        }} = \frac{1}{{2{\rho ^2}}}\left( {tr({S_{{{\bf{x}}_t}}}) + {m_{{{\bf{x}}_t}}}^T{m_{{{\bf{x}}_t}}}} \right).
\end{aligned}
\label{eq22}
\end{equation}
Similarly, for $-{E_{{\bf{H}}\sim q\left( {{\bf{H}}|{{\bf{y}}_t}} \right)}}\left[ {\log p\left( {\bf{H}} \right)} \right]$
\begin{equation}
\begin{aligned}
 - \int_{\bf{H}} {q\left( {{\bf{H}}|{{\bf{y}}_t}} \right)} \log p\left( {\bf{H}} \right)d{\bf{H}} &= \frac{1}{2}{E_{{\bf{H}}\sim q\left( {{\bf{H}}|{{\bf{y}}_t}} \right)}}\left[ {{{\bf{H}}^H}{\bf{H}}} \right]+C_5\\
{\rm{                                     }} &= \frac{1}{2}tr({S_{\bf{H}}}) + \frac{1}{2}{m_{\bf{H}}}^T{m_{\bf{H}}}+C_5.
\end{aligned}
\label{eq23}
\end{equation}
For the decoder term in \eqref{eq17}, the  Monte Carlo method is used to approximate the expectation
\begin{equation}
\begin{aligned}
 &- \int_{{{\bf{x}}_t},{\bf{H}}} {q\left( {{\bf{H}}|{{\bf{y}}_t}} \right)q\left( {{{\bf{x}}_t}|{{\bf{y}}_t}} \right)} \log p\left( {{{\bf{y}}_t}|{\bf{H}},{{\bf{x}}_t}} \right)d{{\bf{x}}_t}d{\bf{H}}\\&
 = {E_{{\bf{H}},{{{\bf{x}}_t}}}}\left[ {{{({{\bf{y}}_t} - {\bf{H}}{{\bf{x}}_t})}^H}({{\bf{y}}_t} - {\bf{H}}{{\bf{x}}_t})} \right]\\&
{\rm{ = }}{E_{{\bf{H}},{{{\bf{x}}_t}}}}\left[ {{{\bf{y}}_t}^H{{\bf{y}}_t} - {{\bf{y}}_t}^H{\bf{H}}{{\bf{x}}_t} - {{\bf{x}}_t}^H{{\bf{H}}^H}{{\bf{y}}_t} - {{({\bf{H}}{{\bf{x}}_t})}^H}({\bf{H}}{{\bf{x}}_t})} \right]\\&
 = {E_{\bf{H}}}\left[ {tr({\bf{H}}{S_{{{\bf{x}}_t}}}{{\bf{H}}^H}) + {{({\bf{H}}{m_{{{\bf{x}}_t}}} - {{\bf{y}}_t})}^H}({\bf{H}}{m_{{{\bf{x}}_t}}} - {{\bf{y}}_t})} \right]\\&
 \approx \frac{1}{L}\sum\limits_{{\widehat{{\bf{H}}}_{\rm{l}}}} {tr({\widehat{{\bf{H}}}_{\rm{l}}}{S_{{{\bf{x}}_t}}}{\widehat{{\bf{H}}}_{\rm{l}}}^H) + {{({\widehat{{\bf{H}}}_{\rm{l}}}{m_{{{\bf{x}}_t}}} - {{\bf{y}}_t})}^H}({\widehat{{\bf{H}}}_{\rm{l}}}{m_{{{\bf{x}}_t}}} - {{\bf{y}}_t})},
\end{aligned}
\label{eq24}
\end{equation}
where the \emph{reparameterization trick}  is used in sampling operation as the  Monte Carlo method is not differentiable which  impediments towards BP operation,  sampling a ${\widehat{{\bf{H}}}_{\rm{l}}}$ from distribution $\mathcal{CN}\left( {{m_{{{\bf{H}}_t}}},{S_{{{\bf{H}}_t}}}} \right)$ is equivalent to sampling a  $\bf{h}$  from distribution $\mathcal{CN}(0,I)$  and let ${\widehat{{\bf{H}}}_{\rm{l}}} = {{m_{{{\bf{H}}_t}}}} + {{S_{{{\bf{H}}_t}}}} \times \bf{h} $, and then the encode network can be trained as the sampling operation does not need to participate in the gradient descent process.

\end{document}